\documentstyle[rotate,psfig,twocolumn]{mn}
\title[Compact steep-spectrum sources]{Compact steep-spectrum sources from the S4 sample}

\author[D.J. Saikia et al.]{D.J. Saikia$^{1}$, S. Jeyakumar $^{1}$
C.J. Salter$^{2}$, P. Thomasson $^{3}$, R.E. Spencer$^{3}$ 
\newauthor and F. Mantovani  $^{4}$ \\
$^{1}$ Tata Institute of Fundamental Research, National Centre for Radio
Astrophysics, P.B. No. 3, Ganeshkhind, Pune 411 007, India \\
$^{2}$ Arecibo Observatory, HC3 Box 53995, Arecibo, Puerto Rico 00612, USA  \\
$^{3}$ University of Manchester, NRAL, Jodrell Bank, Macclesfield, Cheshire, SK11 9DL, UK \\
$^{4}$ Istituto di Radioastronomia del CNR, Via P. Gobetti 101, I-40129
Bologna, Italy \\
}

\date{Received:}
\begin{document}
\maketitle

\begin{abstract}
We present the results of 5-GHz observations with the VLA A-array 
of a sample of candidate Compact Steep Spectrum sources 
(CSSs) selected from the S4 survey. We also estimate the symmetry parameters of  
high-luminosity CSSs selected from different samples of radio sources, and 
compare these with the larger sources of similar luminosity to understand
their evolution and the consistency of the CSSs with the unified scheme for 
radio galaxies  and quasars. The 
majority of CSSs are likely to be young sources advancing outwards through a 
dense asymmetric environment.  The radio properties of CSSs are found to be
consistent with the unified scheme, in which the axes of the quasars are 
observed close to the line of sight, while radio galaxies are observed 
close to the plane of the sky.
\end{abstract}


\section{Introduction}
High-resolution radio observations of compact steep-spectrum sources (CSSs) and 
the GHz-peaked spectrum (GPS) sources 
(cf. Fanti et al. 1990; Sanghera et al. 1995 and references therein; Dallacasa
et al. 1995, 1998; Stanghellini et al. 1997; O'Dea 1998) reveal
a variety of structures reminiscent of those seen in the more extended 
sources. For the present study, a CSS source is defined as being less than
about 20 kpc in size ($H_o =100$ km s$^{-1}$ Mpc$^{-1}$ and $q_0 = 0$), and 
having a steep high-frequency radio spectrum ($\alpha \geq 0.5$,
where S$\propto \nu^{-\alpha}$). The highly compact
GPS sources, with turnover frequencies in the GHz range, are included.
The majority of CSS and GPS sources have double-lobed structures, often showing
nuclear or core components. The GPS sources have been proposed to be miniature
versions of the classical Fanaroff-Riley class II sources, and have been 
proposed to evolve from a GPS to a CSS and then on to a larger FRII source 
(cf. Carvalho 1985; Mutel \& Phillips 1988; Fanti et al. 1995; Begelman 1996;
Readhead et al. 1996a,b; Owsianik \& Conway 1998; de Vries et al. 1998).

The detection of radio cores in a large number of the GPS and CSS 
sources has made it possible to estimate the symmetry parameters of this class 
of objects reliably, and permits the probing of their environment on these scales.
Testing of the consistency of the CSSs with the expectations of the unification 
scheme (Scheuer 1987; Barthel 1989; Antonucci 1993; Urry \& Padovani 1995) for 
radio galaxies and quasars is also possible.
In an earlier study, Saikia et al. (1995, hereinafter referred to as S95) found
CSSs to be more asymmetric than the larger sources suggesting that their jets 
on these scales are propagating 
outwards through an asymmetric environment. They also reported
that the core strengths and symmetry parameters of CSSs were consistent with
the unified scheme.

In this paper, we first present radio observations with the Very Large Array
(VLA) of a sample of CSSs selected from the S4 survey at 5 GHz (Pauliny-Toth et al. 
1978), and which were earlier found to be unresolved or partially resolved with
the Westerbork telescope at 5 GHz (Kapahi 1981). We then re-examine the trends
reported in S95 using the S4 sample, as well as more recent information 
on the detection of radio cores in the samples considered by S95. 
The radio observations  are presented in Section 2, while in Section 3
we describe the sample of sources used in our estimates of the symmetry parameters.
In Section 4, we examine the dependence of the symmetry parameters on 
linear size, optical identification
and core strength, with emphasis on studying the 
evolution and consistency with the unified scheme of CSSs. 
The results are summarized in Section 5.

\section{The S4 sample of radio sources}
Complete samples of radio sources provide a powerful means of studying the 
properties and evolution of radio sources with age and cosmic epoch, as well
as selecting individual sources for detailed studies of the physical processes
responsible for their energetics and structure. However, only a few complete
samples of radio sources have reasonably satisfactory information on the optical
identification of the host galaxies and measurements of their redshifts. These
include the 3CR (Laing, Riley \& Longair 1983) and MRC (Large et al. 1981; 
Kapahi et al. 1998a, b) samples selected at low radio frequencies, the 2-Jy all-sky
survey at 2.7 GHz (Wall \& Peacock 1985, and references therein), the PKS 100-mJy 
deep survey (Dunlop et 
al. 1989) and the S4 survey (Pauliny-Toth et al. 1978; Stickel \& K\"{u}hr 1994).
The S4 survey is complete to 0.5 Jy at 5 GHz and lies intermediate in brightness
between the Wall \& Peacock (1985) and the Dunlop et al. (1989) surveys. It
covers the region between $+$35$^\circ\leq\delta\leq+70^\circ$ and consists of 270
sources with $\mid b^{II}\mid \geq 10^\circ$. This complete sample, with 
updates of the optical 
identifications and redshifts, has been presented by Stickel \& K\"{u}hr (1994).

\begin{table}
\caption{Observed properties of the sources 
}
\begin{tabular}{c l rrr r r r  r}
Source  & Opt & \multicolumn{3}{c}{Beam size} & $\sigma$ & S$_{c}$  & S$_{tot}$  \\

& &  maj. & min. & PA             &          &            &             \\
(1) & (2) &  \multicolumn{3}{c}{(3)} & (4) & (5) & (6)  \\
\\[1mm]
0011+34  & EF & 0.51 & 0.37 & 150 & 0.06 &      &  395           \\
0223+34  & Q  & 0.47 & 0.36 & 150 & 0.13 &      & 1638          \\
0307+44  & Q  & 0.48 & 0.34 & 151 & 0.06 &  190 &  600          \\
0420+34  & EF & 0.47 & 0.37 & 152 & 0.06 &  257 &  476          \\
0625+50  &    & 0.51 & 0.34 &   1 & 0.06 &      &  430           \\
0639+59  & EF & 0.54 & 0.35 &   8 & 0.10 &      &  526           \\
0655+69  & EF & 0.60 & 0.35 &  10 & 0.08 &      &  689           \\
0657+68  & G  & 0.56 & 0.33 &   9 & 0.10 &   59 &  499           \\
0659+44  & EF & 0.48 & 0.34 &   6 & 0.07 &      &  560           \\
0703+46  & Q  & 0.44 & 0.33 &   7 & 0.06 &      &  594           \\
0707+68  & Q  & 0.58 & 0.35 &   5 & 0.11 &      &  745           \\
0809+48  & Q  & 0.48 & 0.34 &   5 & 0.33 &      & 3967           \\
0827+37  & Q  & 0.46 & 0.35 & 177 & 0.11 &      &  847           \\
0840+42  & EF & 0.48 & 0.35 &   0 & 0.05 &      &  554           \\
0902+49  & Q  & 0.50 & 0.34 &   8 & 0.05 &  401 &  466           \\
0906+43  & Q  & 0.49 & 0.34 &   6 & 0.11 &  871 & 1570           \\
0945+66  & G  & 0.53 & 0.37 &  13 & 0.10 &      & 1215           \\
1014+39  & G  & 0.44 & 0.37 &  10 & 0.06 &      &  487           \\
1020+59  & Q  & 0.47 & 0.37 &  14 & 0.05 &   22 &  484           \\
1128+45  & Q  & 0.46 & 0.37 &  37 & 0.06 &      &  656           \\
1131+43  & G  & 0.45 & 0.37 &  39 & 0.04 &      &  500           \\
1133+43  & EF & 0.45 & 0.37 &  30 & 0.05 &      &  505           \\
1138+59  & EF & 0.49 & 0.36 &  30 & 0.05 &      &  772           \\
1153+59  & G  & 0.50 & 0.33 &  12 & 0.09 &   80 &  416           \\
1156+54  & EF & 0.51 & 0.34 &  10 & 0.08 &      &  625           \\
1225+36  & Q  & 0.50 & 0.36 &  10 & 0.07 &      &  783           \\
1242+41  & Q  & 0.49 & 0.36 &   3 & 0.07 &      &  690           \\
1244+49  & G  & 0.48 & 0.33 &   4 & 0.05 &   40 &  573           \\
1402+66  & EF & 0.55 & 0.33 &  11 & 0.07 &      &  721           \\
1409+52  & G  & 0.50 & 0.34 &  11 & 0.36 &      & 6438           \\
1413+34  & EF & 0.46 & 0.36 &   6 & 0.08 &      &  930           \\
1818+35  & Q  & 0.75 & 0.48 &  72 & 0.17 &   53 &  474           \\
1819+39  & G  & 0.74 & 0.40 &  89 & 0.49 &      &  953           \\
1843+40  & G  & 0.74 & 0.40 &  86 & 0.32 &      &  567           \\
1943+54  & G  & 0.64 & 0.36 & 109 & 0.15 &      &  885           \\
2015+65  & Q  & 0.68 & 0.34 & 112 & 0.17 &      &  575           \\
2207+37  & Q  & 0.51 & 0.38 & 148 & 0.11 &      &  829           \\
2255+41  & Q  & 0.49 & 0.38 & 148 & 0.18 &      & 1063           \\
2304+37  & G  & 0.47 & 0.34 & 155 & 0.07 &      &  570           \\
2311+46  & Q  & 0.51 & 0.34 & 155 & 0.13 &   75 &  760           \\
2323+43  & G  & 0.47 & 0.34 & 156 & 0.08 &      &  892           \\
2358+40  & EF & 0.48 & 0.36 & 153 & 0.10 &      &  581           \\
\\
\end{tabular}
\end{table}

\subsection{Radio observations}
A sample of 42 candidate CSSs from the S4 survey with a flux density at 5 GHz
$\geq$0.5 Jy, and which were either unresolved or
partially resolved with the WSRT at 5 GHz, were observed with the VLA A-array
at 4835 MHz on 1985 February 10. The radio source 0625+50 with S(5 GHz) $=$ 0.493
Jy (Pauliny-Toth et al. 1978) was also observed. Each source was observed in the 
snapshot mode for about 5-10 min. The data were calibrated at the VLA and reduced 
using AIPS.  All flux densities are on the Baars et al. (1977) scale.

The results of the VLA observations are presented as radio images in Figure 1
for all the well-resolved objects, except 0707+68, 1244+49, 1402+66 and
2323+43, which were presented in Sanghera et al. (1995), and the well-known
3CR sources 0809+48 (3C196) and 0906+43 (3C216) whose images are available in the literature.
Some of the observational parameters are summarized in Table 1, which is arranged
as follows. Column 1: source name in the IAU format; column 2: optical identification
where G denotes a galaxy, Q a quasar and EF an empty field; column 3: the major and minor
axes of the restoring beam in arcsec, and its position angle in deg; column 4:
the rms noise in the total-intensity images in units of mJy/beam; columns 5 and 6:
the core and total flux density in the images. The core flux density is the peak
brightness of this feature in units of mJy/beam, while the total flux density in units of 
mJy has been estimated by specifying a box around the radio source. 

\begin{figure*}
\hbox{
  \vbox{
  \psfig{file=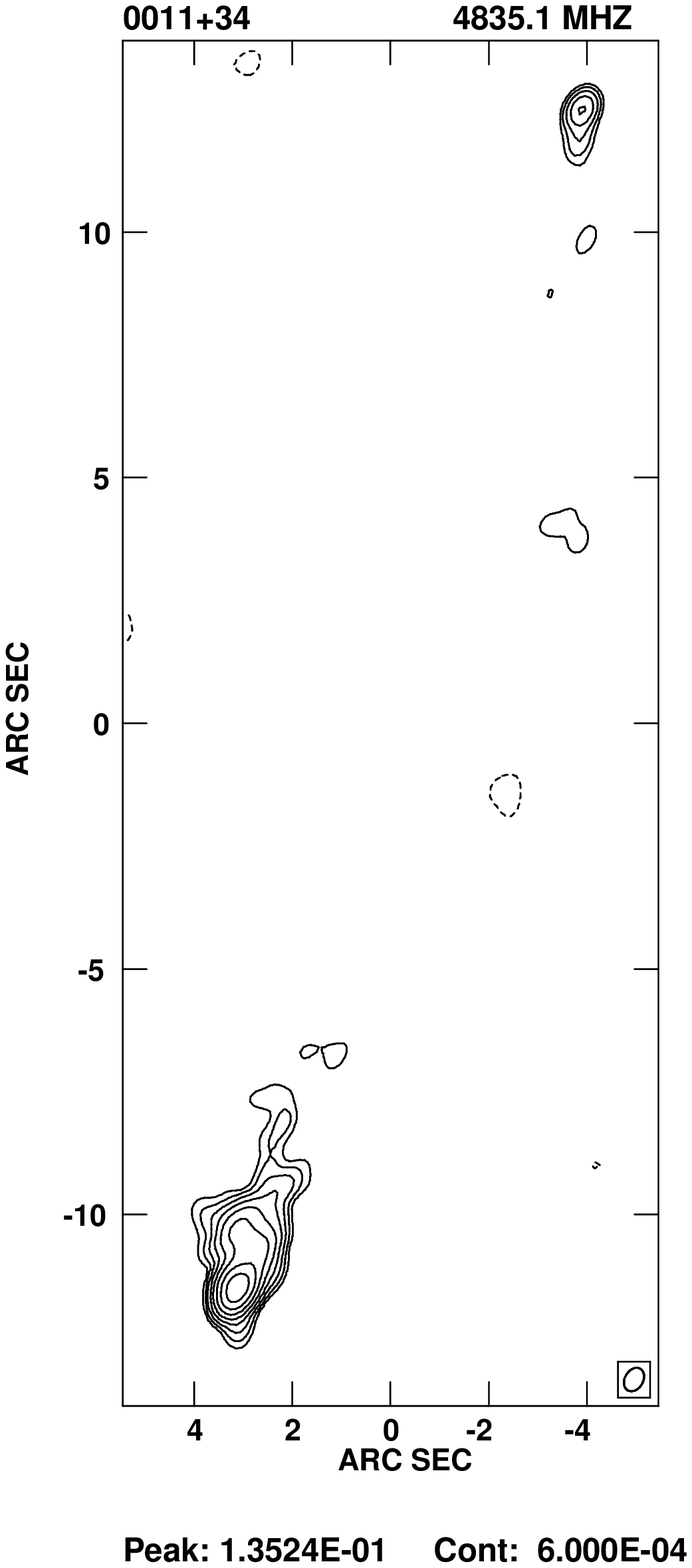,width=2.15in}
  \psfig{file=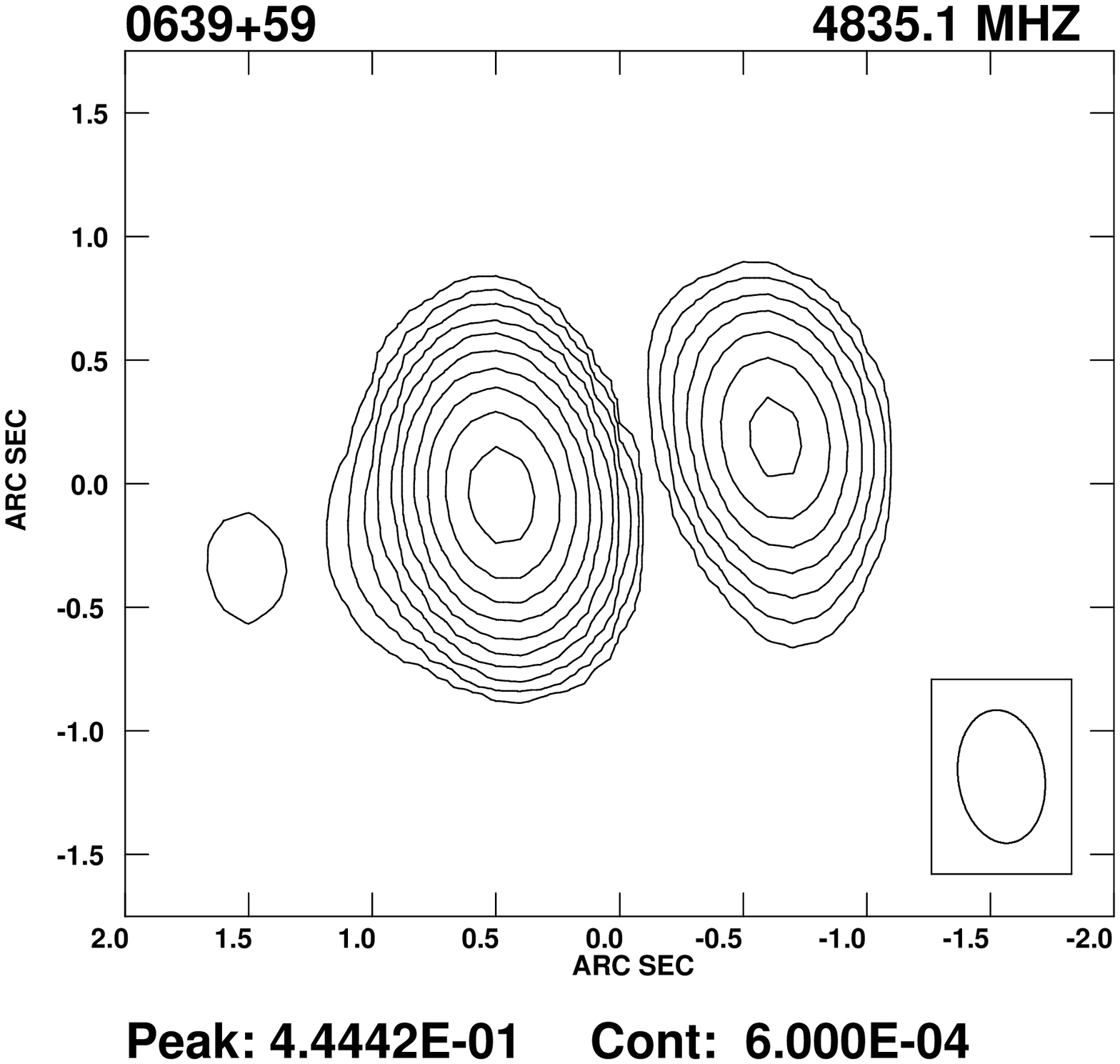,width=2.1in,angle=-90,bbllx=49pt,bblly=125pt,bburx=556pt,bbury=661pt,clip=}
  }
  \vbox{
  \vspace{-0.5cm}
  \psfig{file=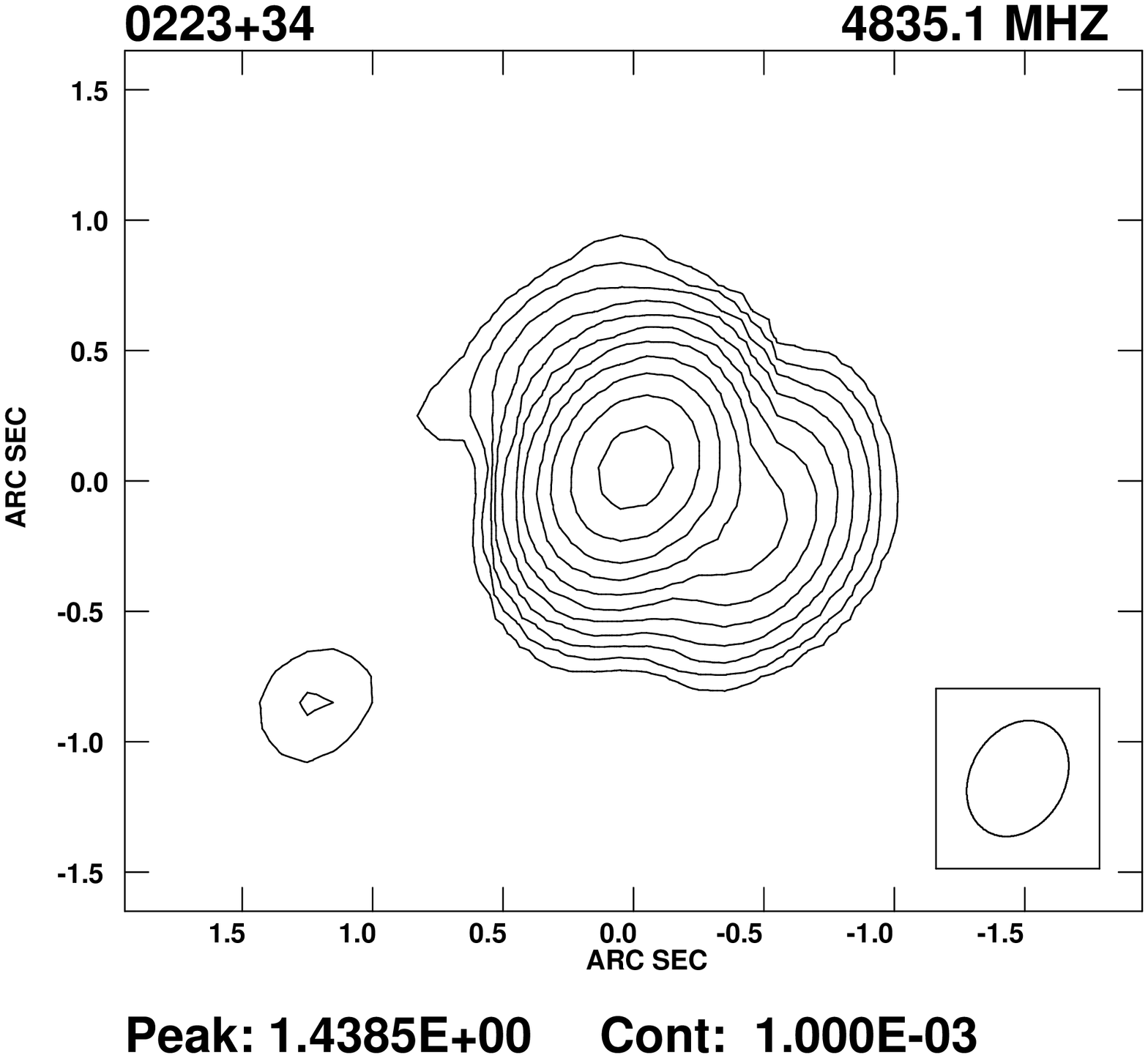,width=2.1in,angle=-90}
  \psfig{file=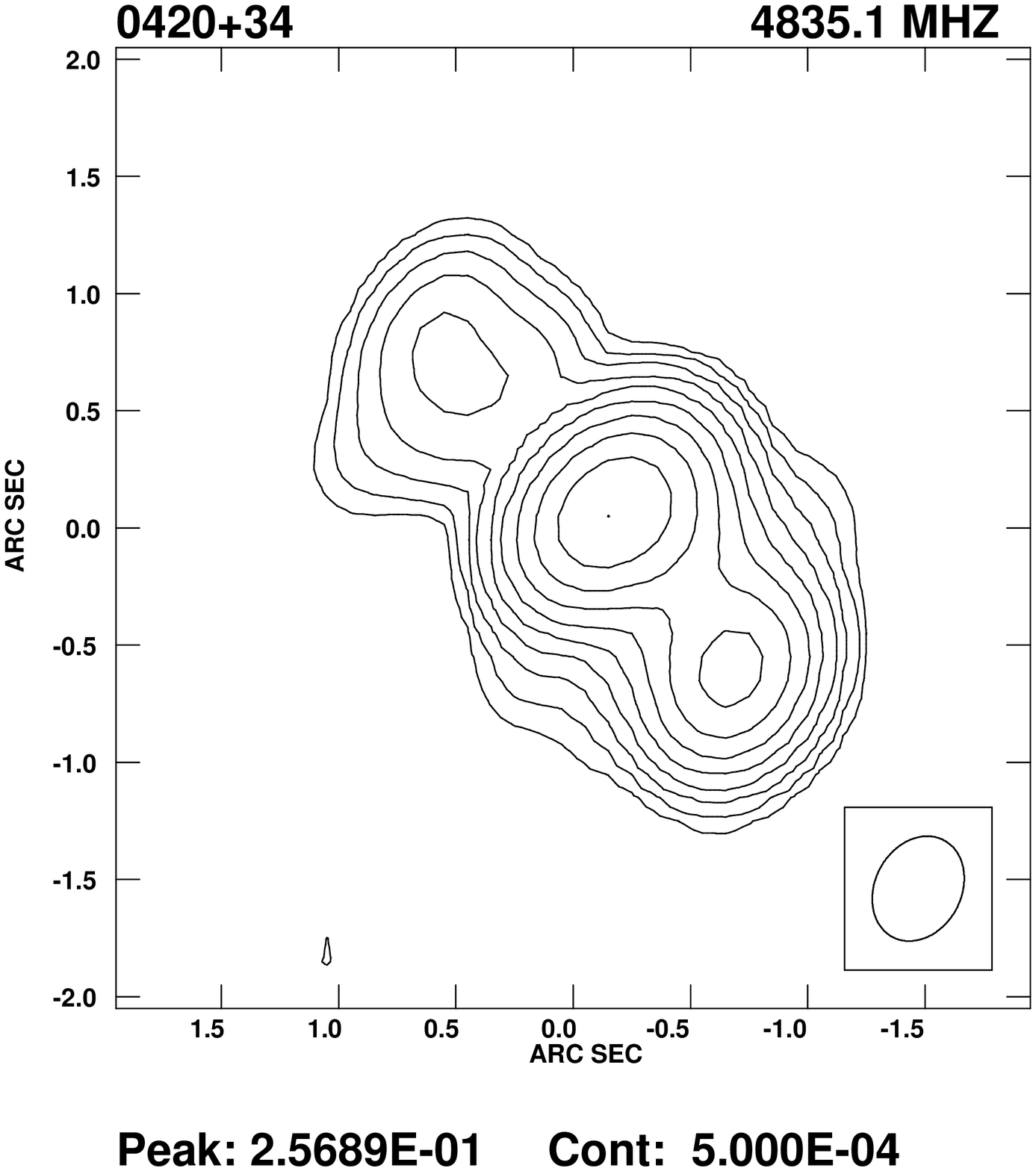,width=2.1in}
  \psfig{file=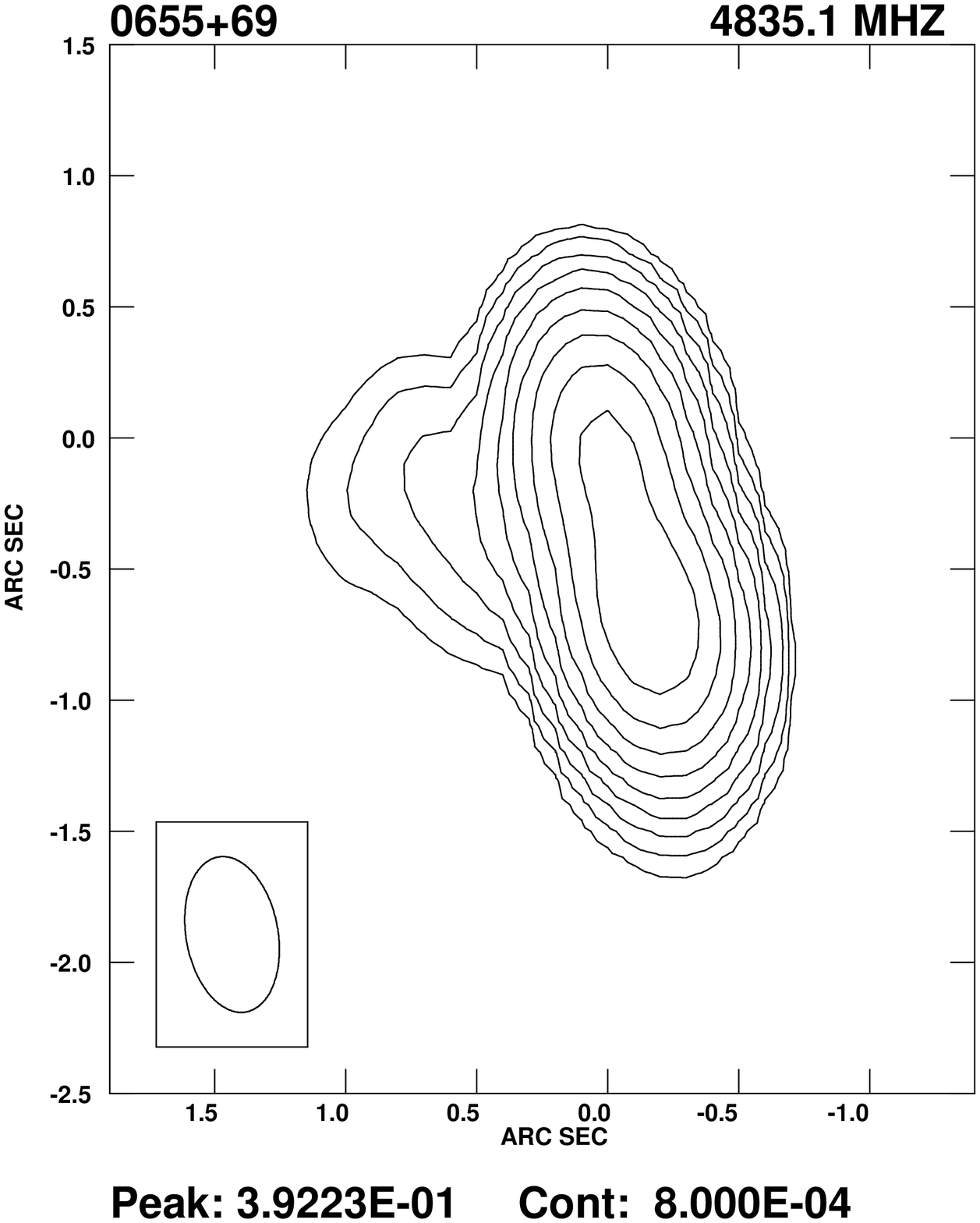,width=2.1in}
  }
  \vbox{
  \psfig{file=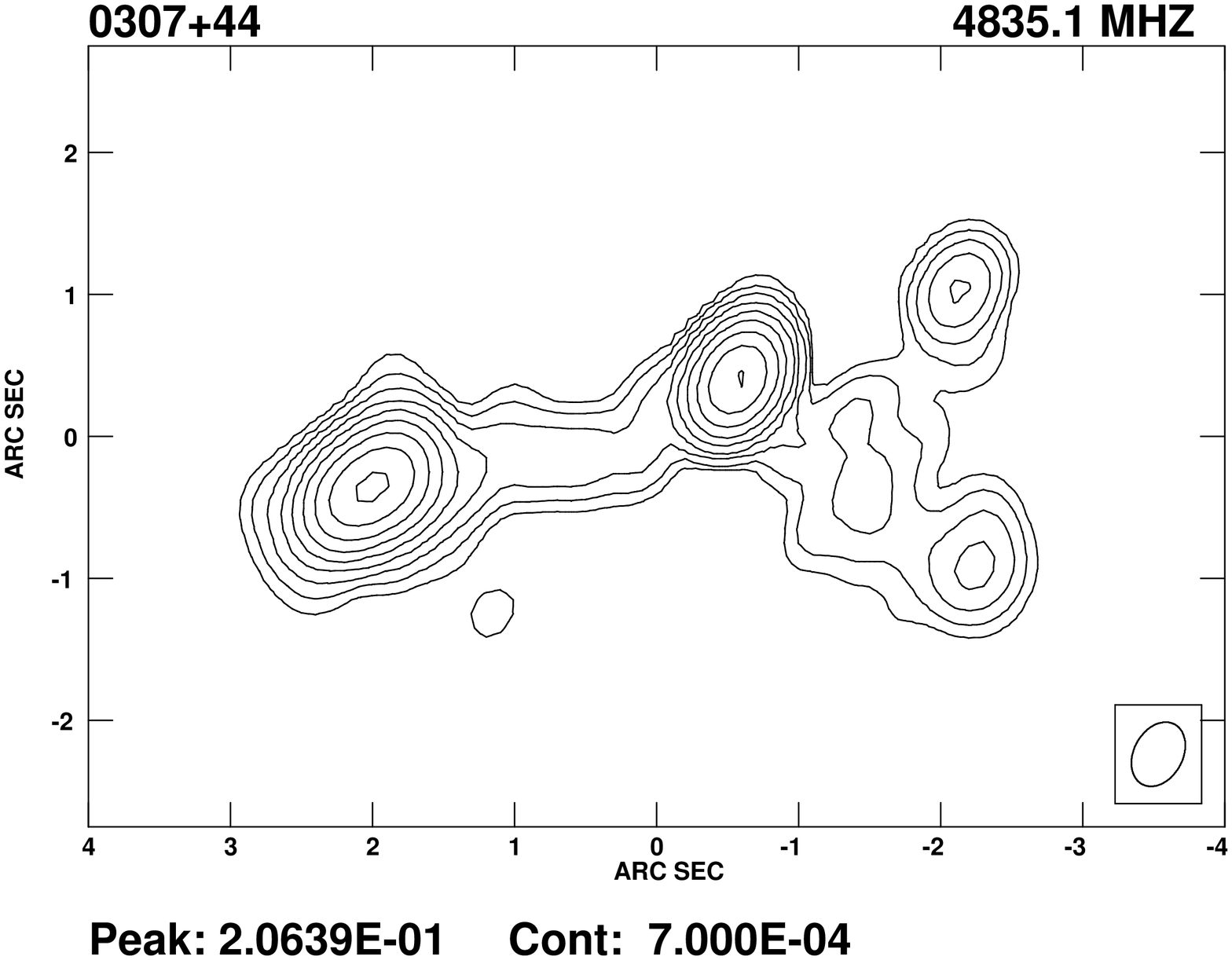,width=2.1in,angle=-90}
  \psfig{file=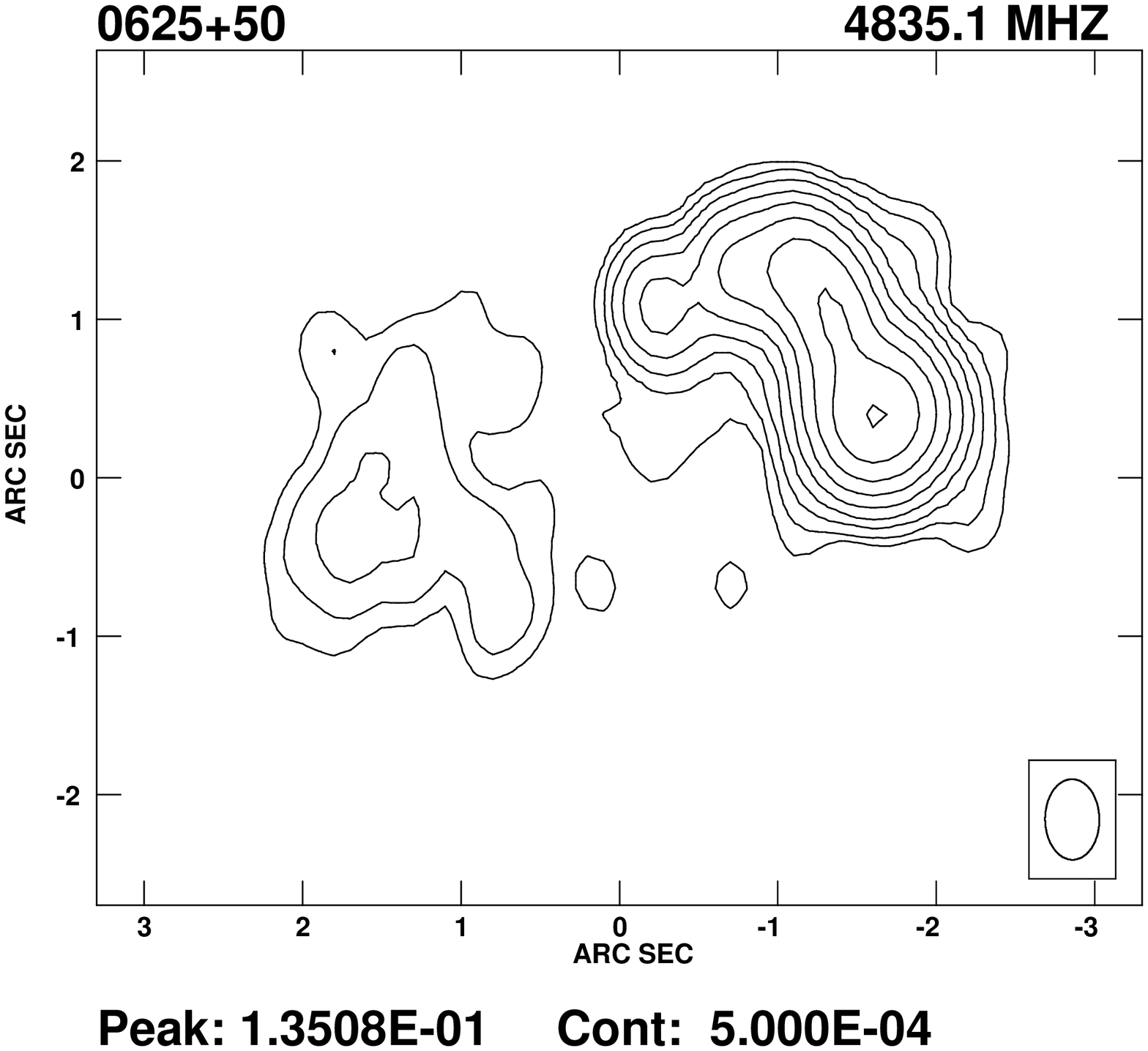,width=2.1in,angle=-90}
  \psfig{file=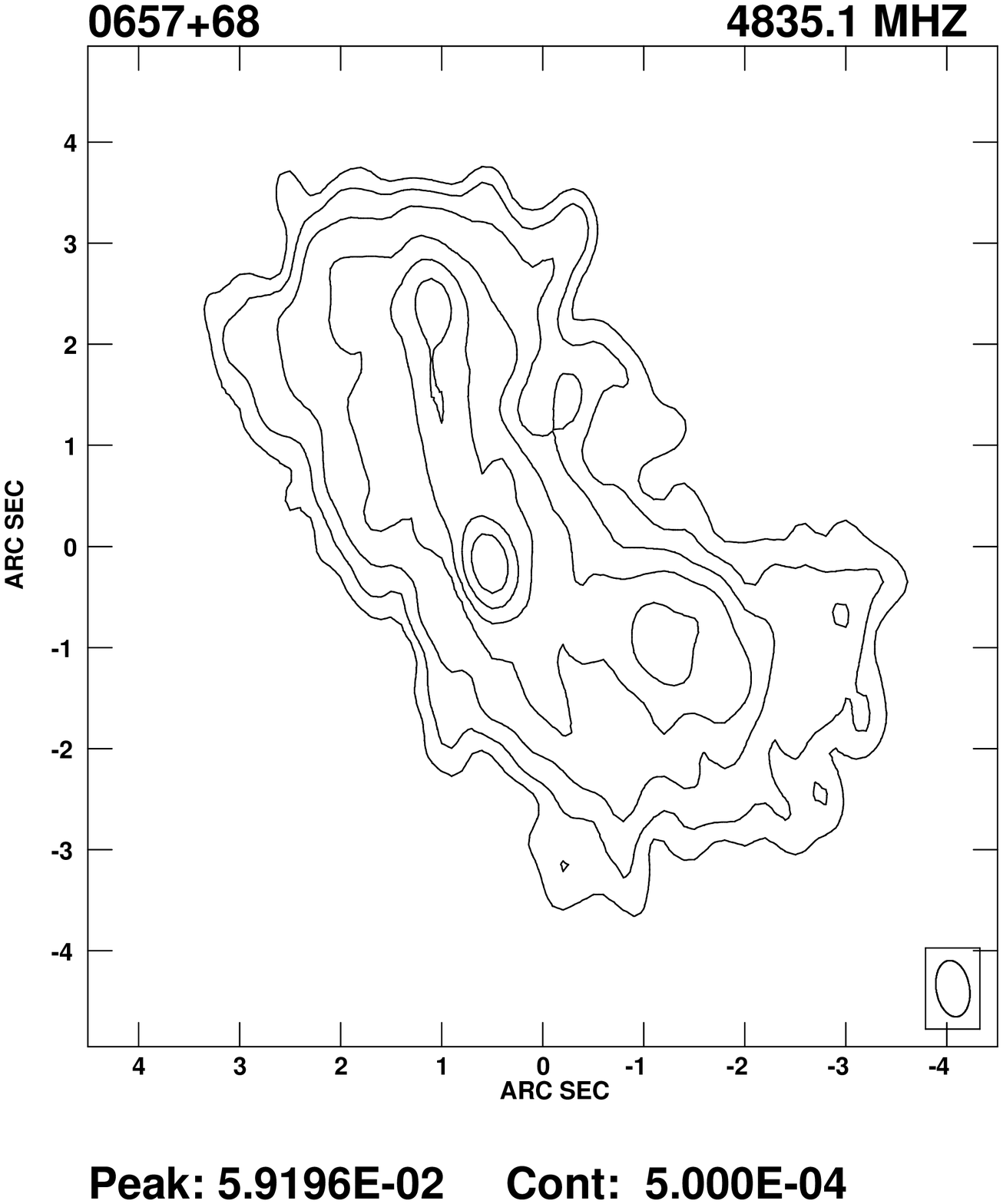,width=2.15in}
  }
}
\caption[VLA images of our sample of S4 sources]{VLA images of our sample of S4 sources. The contour levels for all the images are -1, 1, 2, 4, 8, 16, \ldots times the first contour level. The peak brightness in the image and the level of the first contour in units of mJy/beam are given below the images.
\label{uni:fig:s4images}
}
\end{figure*}

\begin{figure*}
\hbox{
 \vbox{
   \hbox{
   \psfig{bbllx=33pt,bblly=66pt,bburx=571pt,bbury=735pt,clip=,file=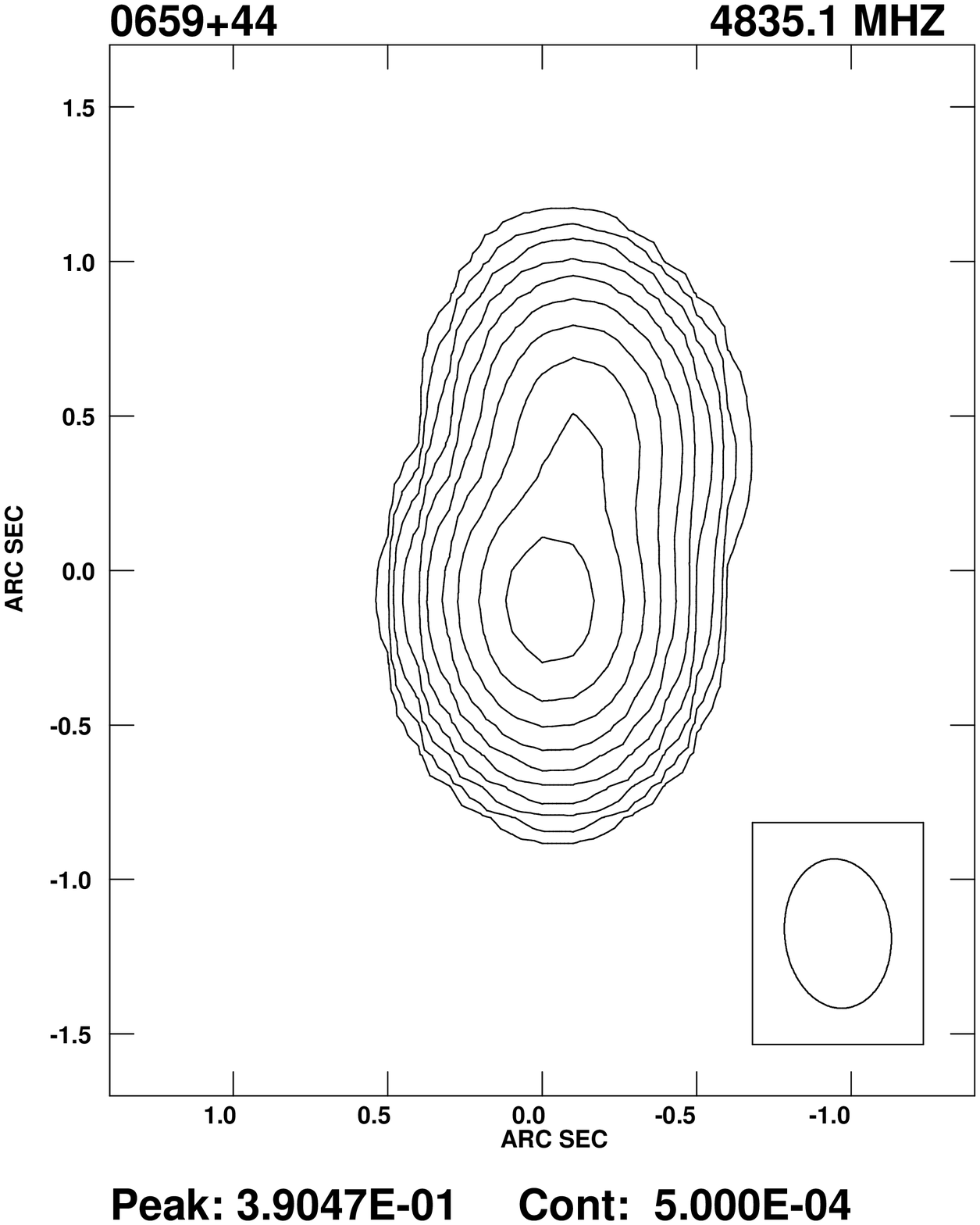,width=2.15in}
   }
   \hbox{
   \psfig{bbllx=35pt,bblly=102pt,bburx=570pt,bbury=699pt,clip=,file=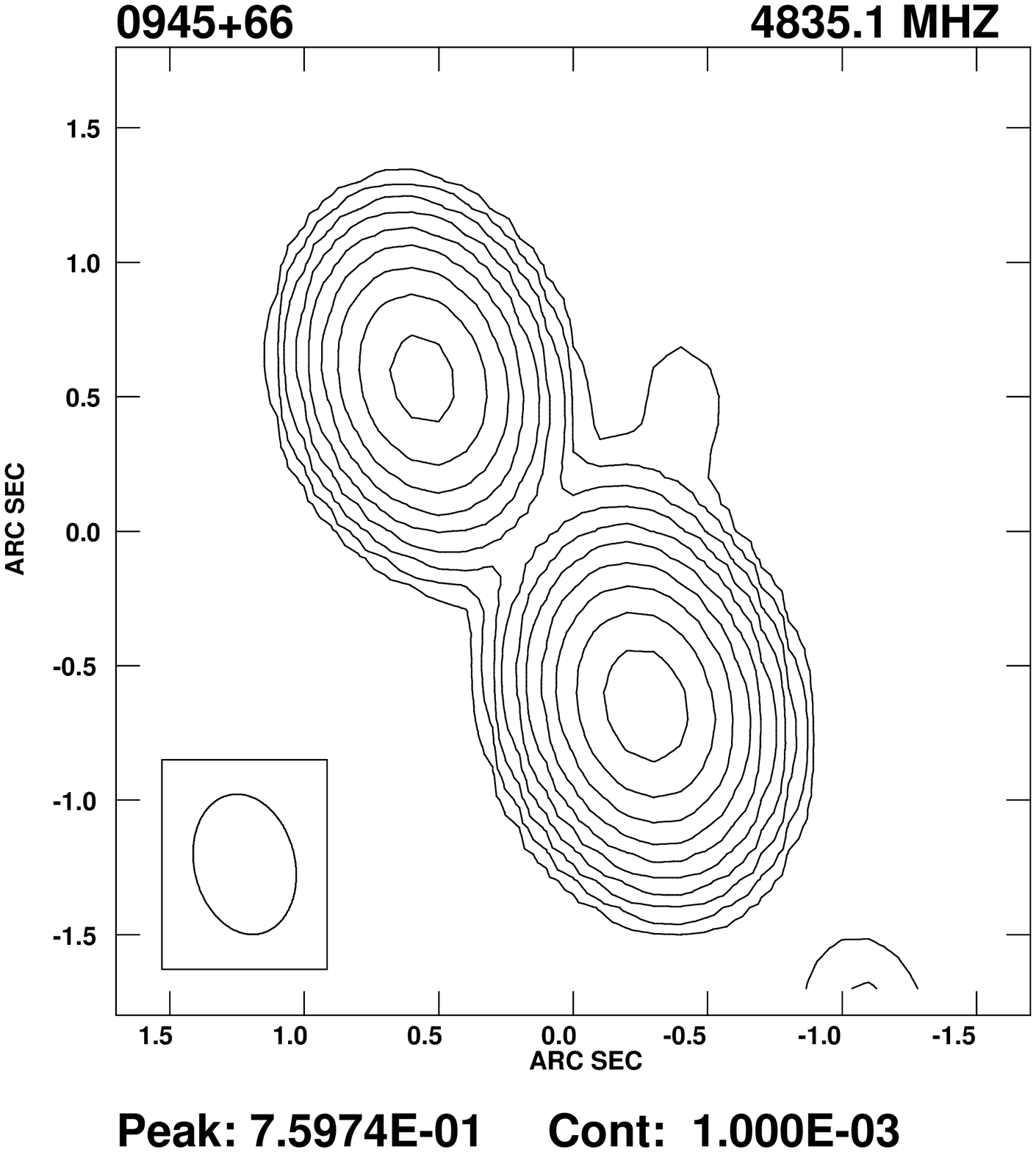,width=2.15in}
   }
   \hbox{
 \psfig{bbllx=35pt,bblly=121pt,bburx=570pt,bbury=683pt,clip=,file=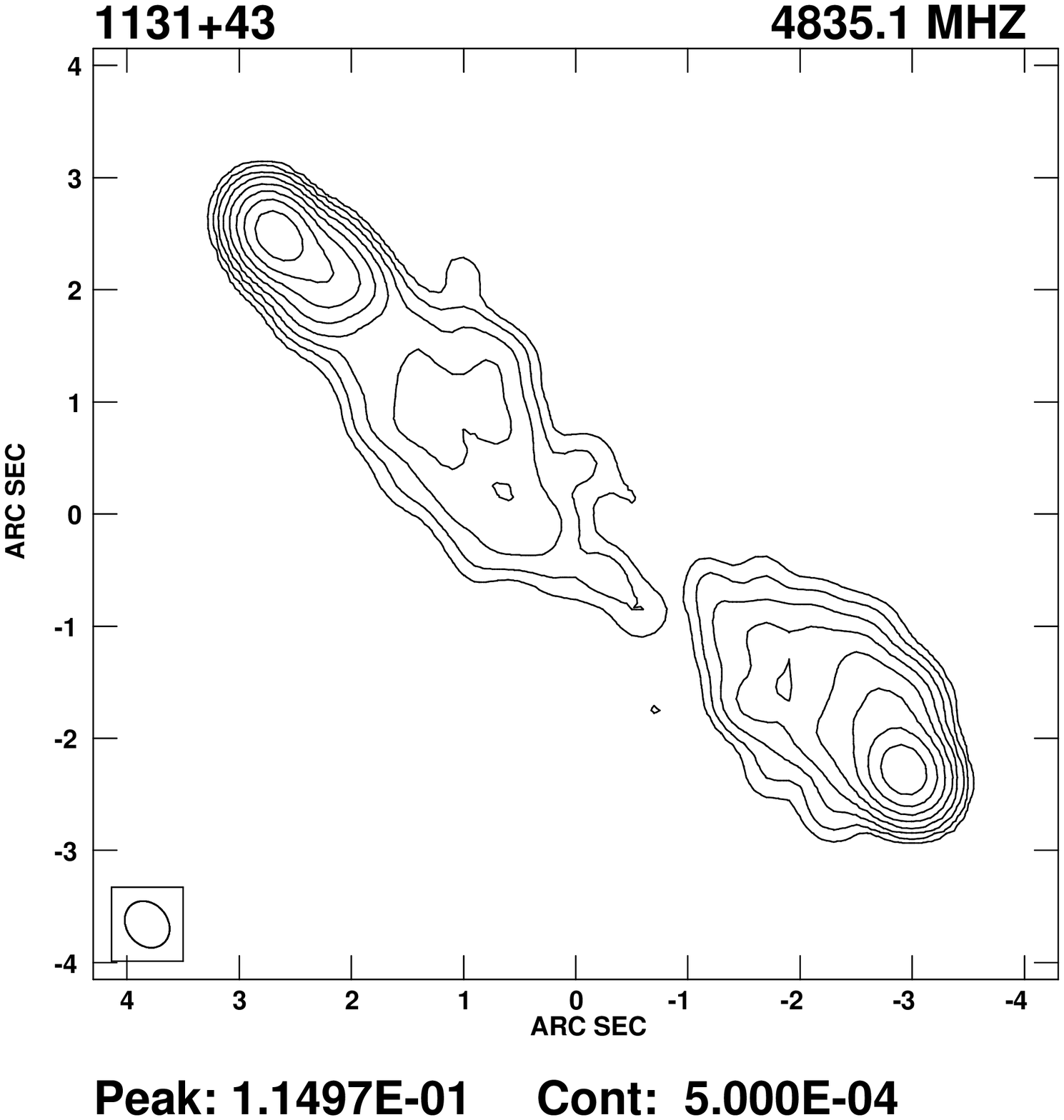,width=2.15in}
   }
 }
 \hbox{
 \vbox{
  \psfig{file=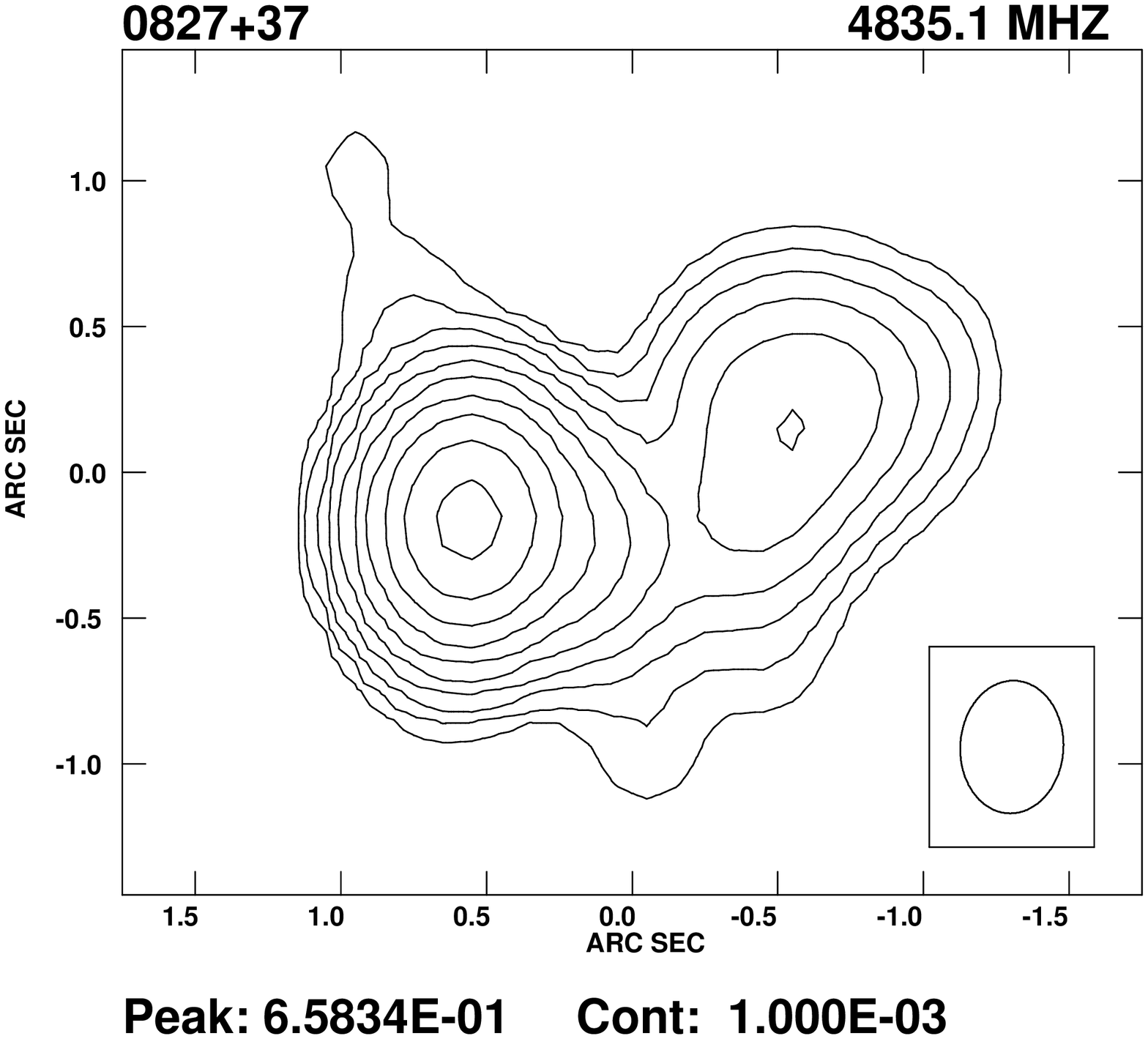,width=2.15in,angle=-90,bbllx=47pt,bblly=113pt,bburx=555pt,bbury=674pt,clip=}
  \psfig{file=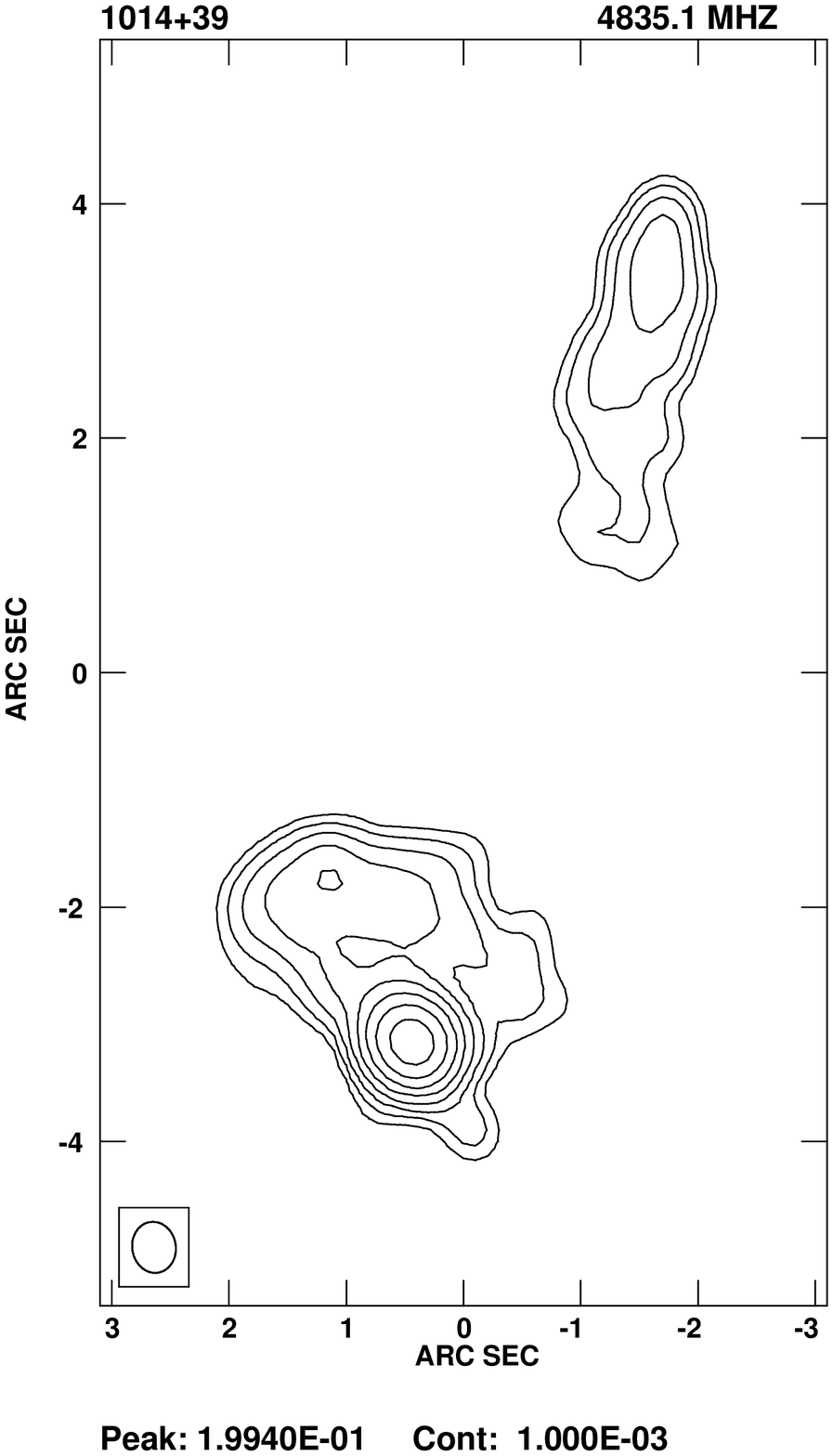,width=2.15in,bbllx=108pt,bblly=56pt,bburx=499pt,bbury=741pt,clip=}
  \psfig{file=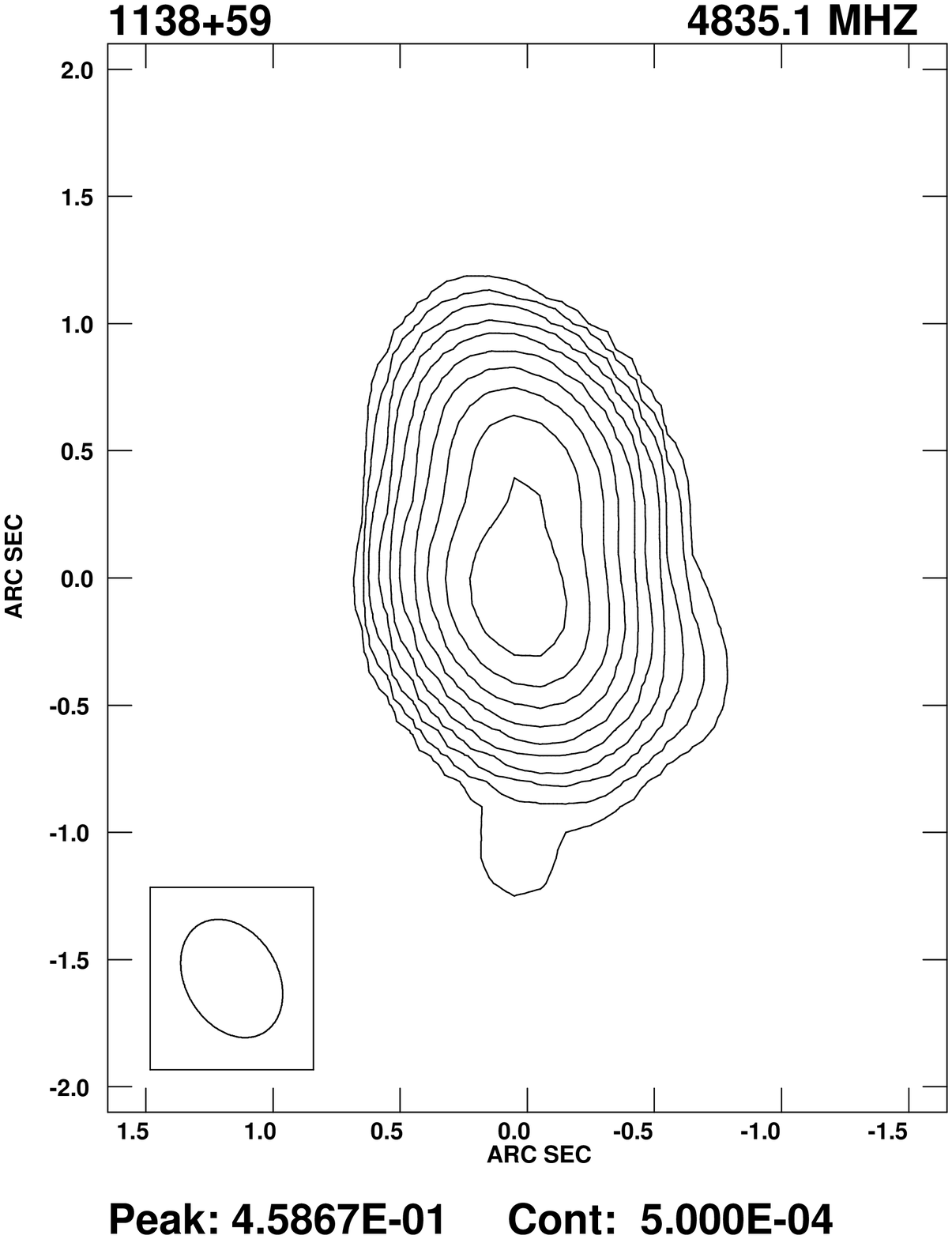,width=2.15in,bbllx=37pt,bblly=54pt,bburx=568pt,bbury=746pt,clip=}
 }
 }

 \vbox{
  \psfig{file=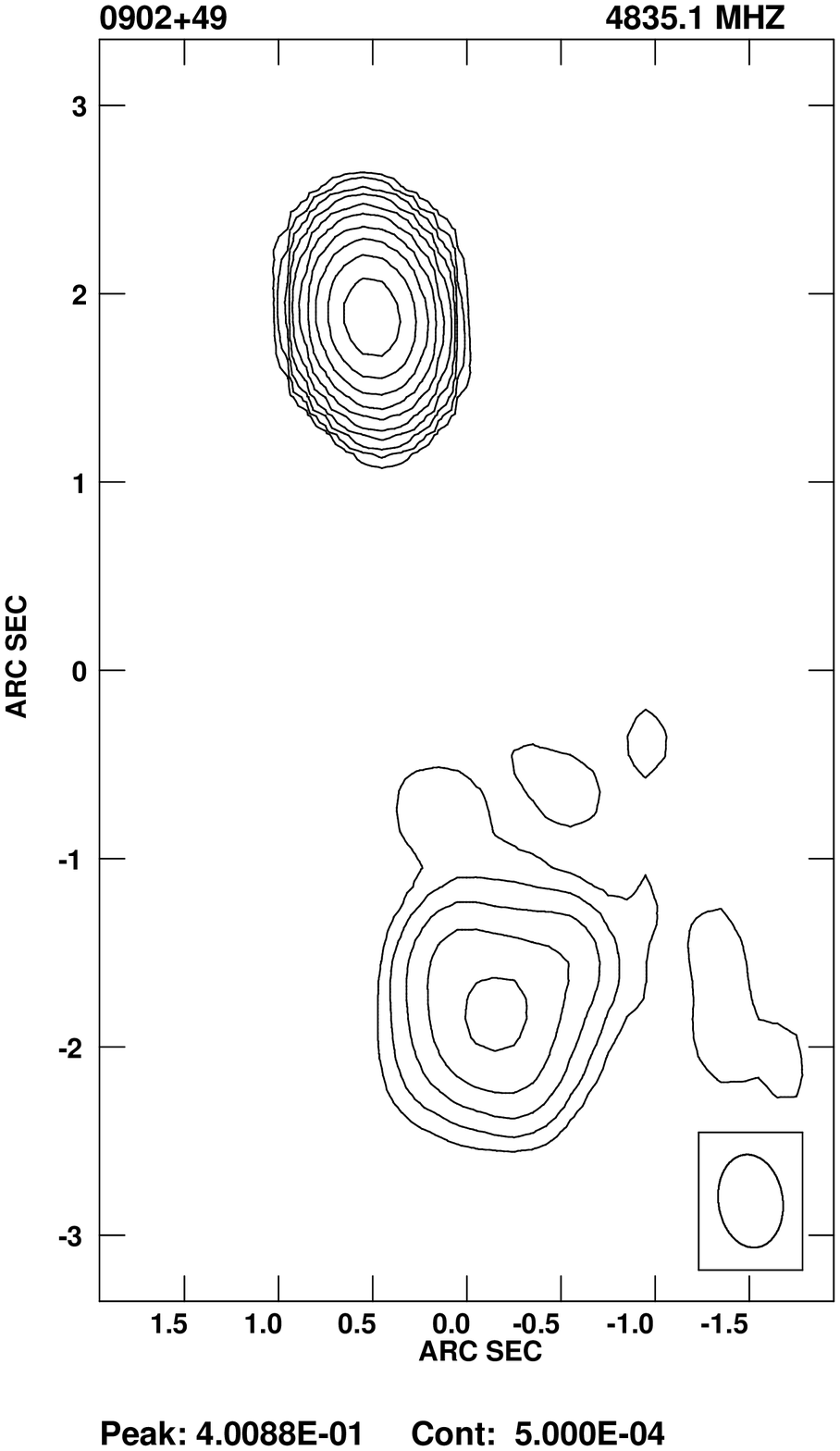,width=2.15in,bbllx=106pt,bblly=58pt,bburx=501pt,bbury=738pt,clip=}
  \psfig{file=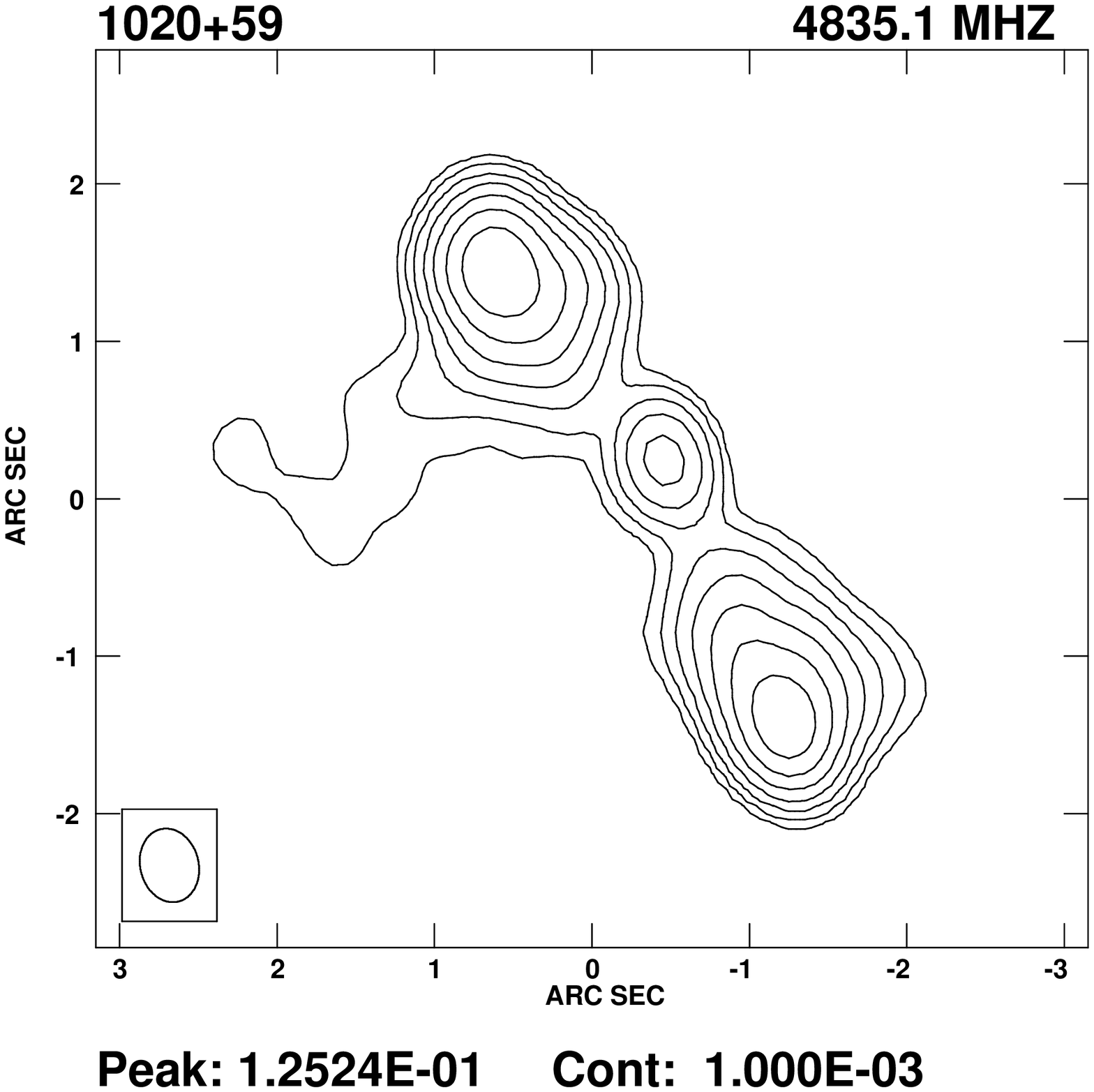,width=2.15in,bbllx=36pt,bblly=135pt,bburx=570pt,bbury=669pt,clip=}
  \psfig{file=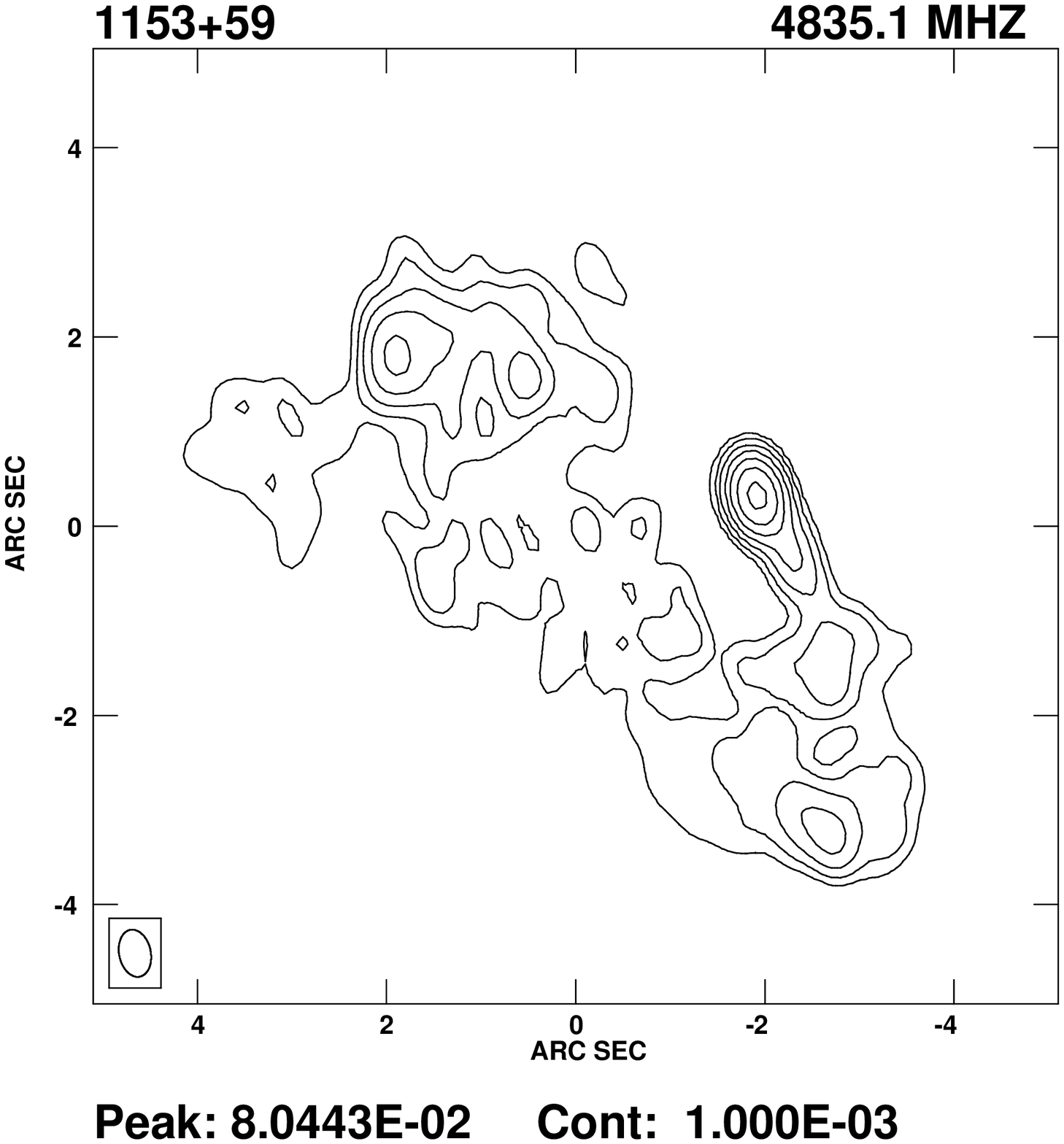,width=2.15in,bbllx=36pt,bblly=113pt,bburx=570pt,bbury=689pt,clip=}
 }
}

\end{figure*}

\begin{figure*}
 \hbox{
 \psfig{file=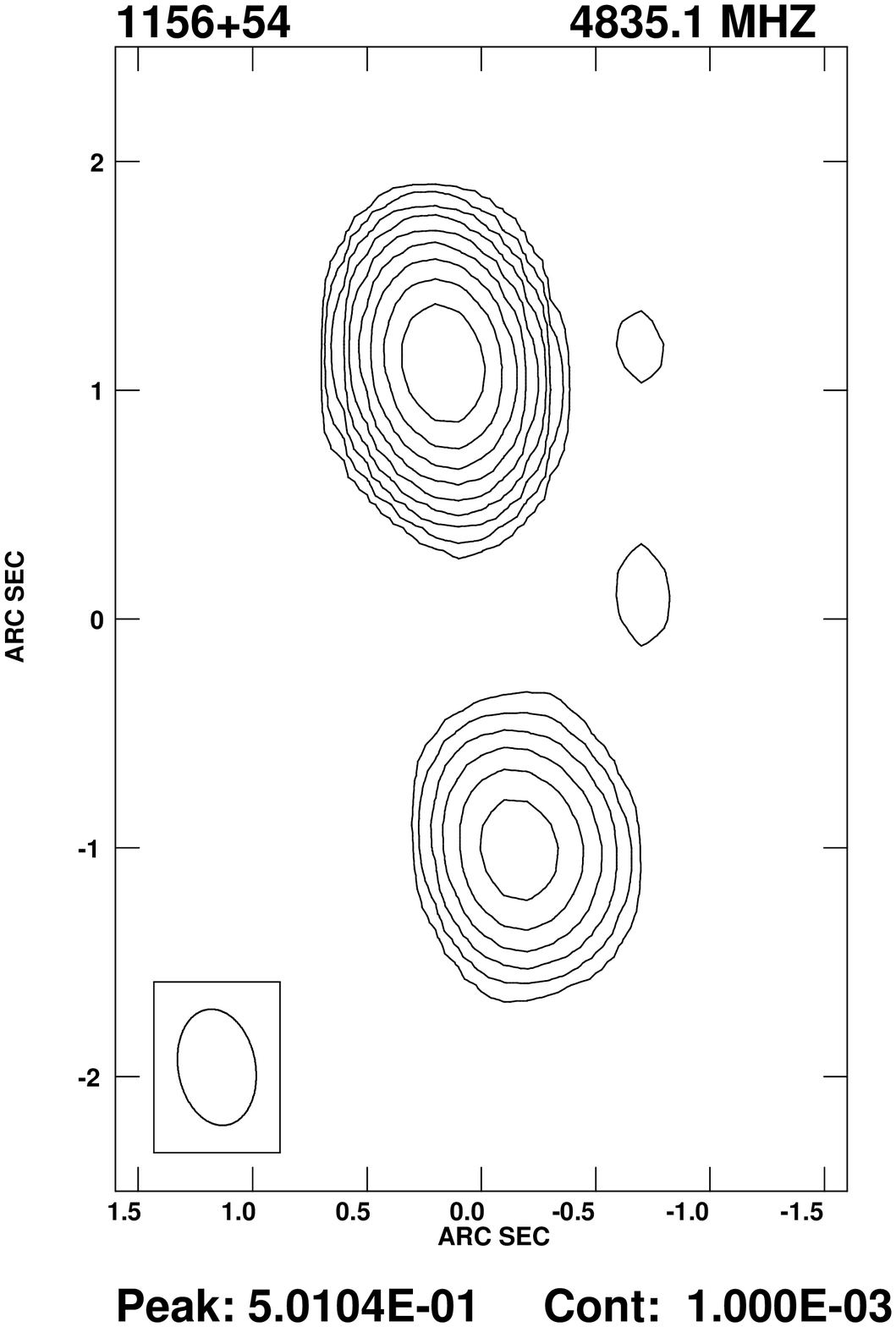,width=2.15in}
  \vbox{
  \psfig{file=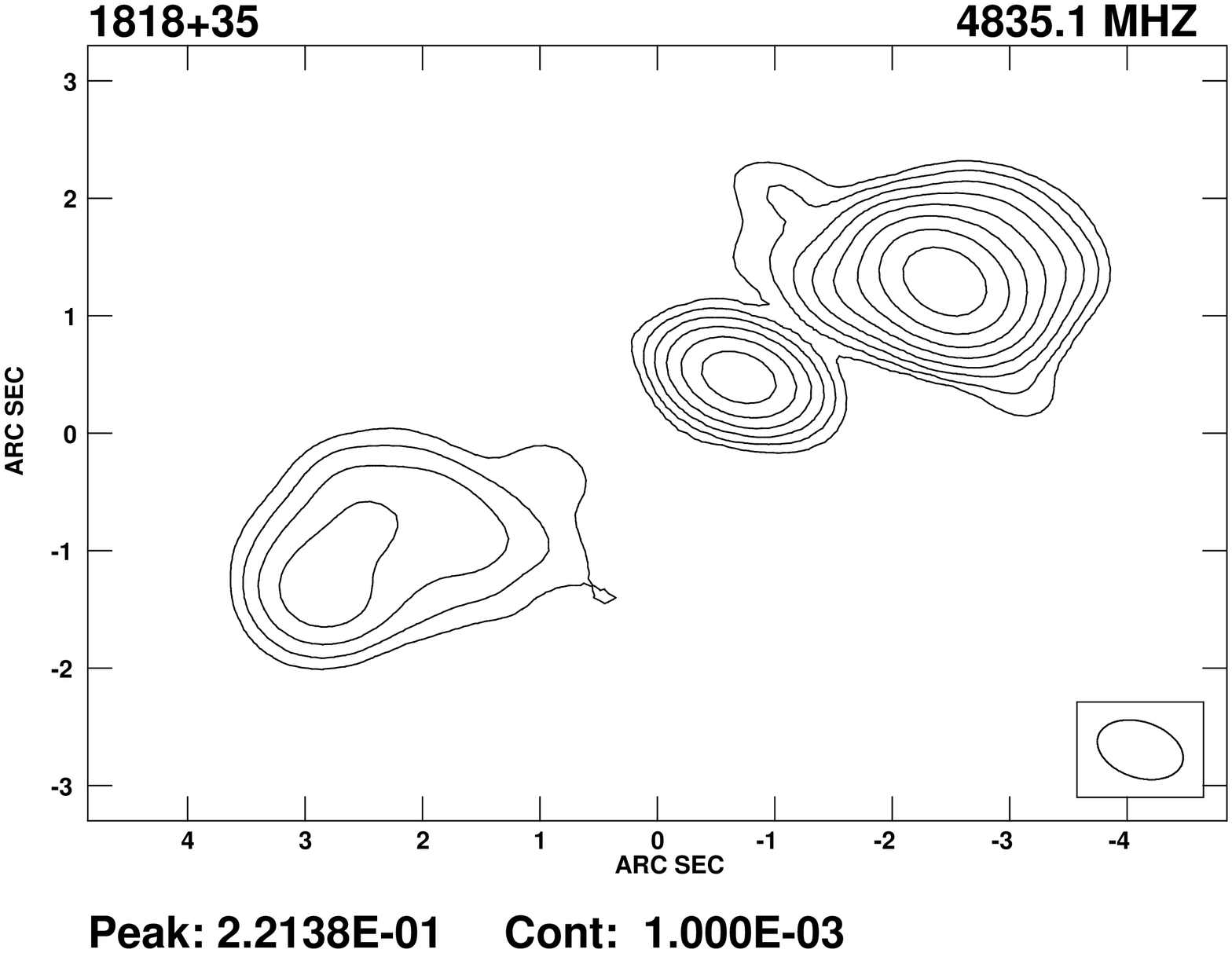,width=2.15in,angle=-90}
  \psfig{file=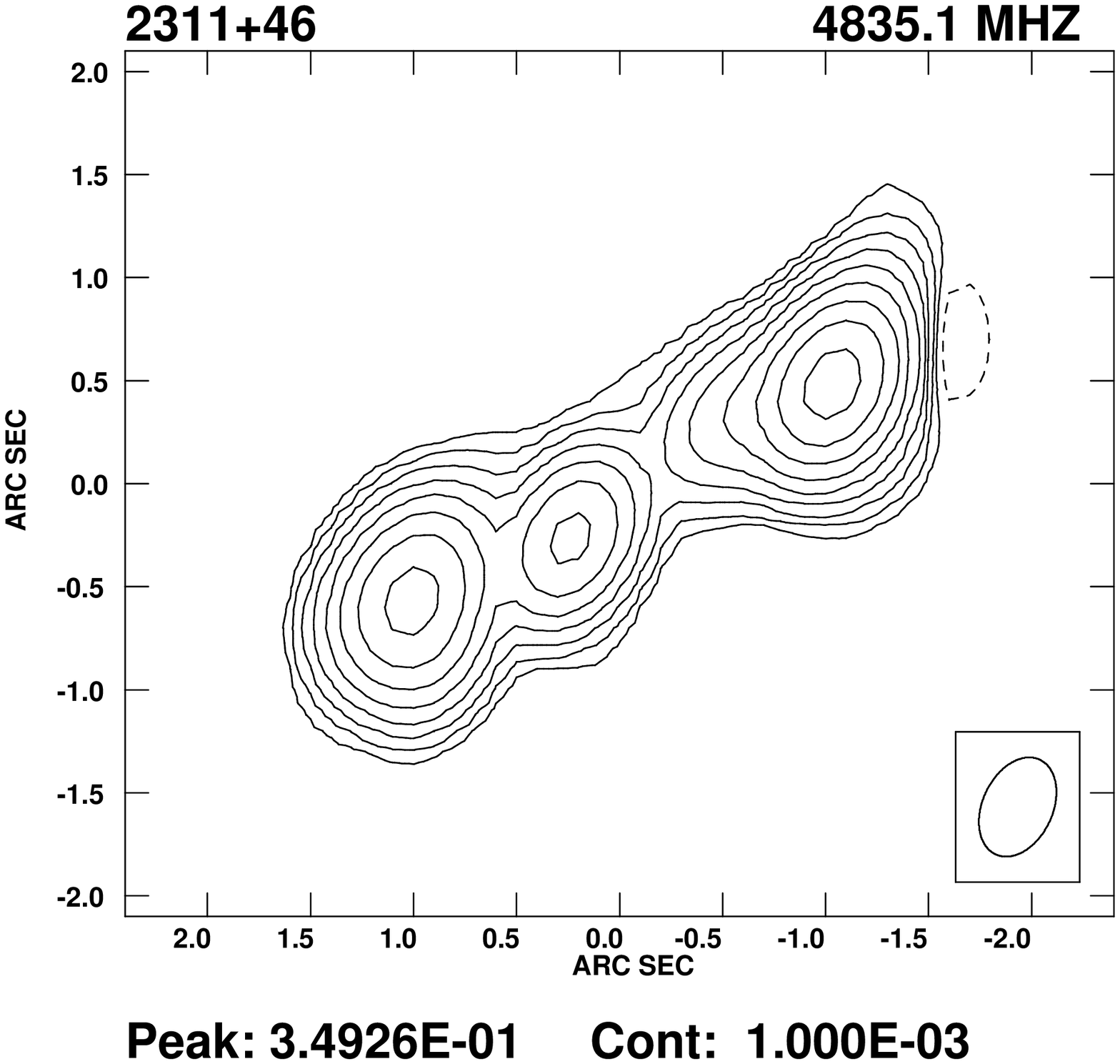,width=2.15in,angle=-90}
   }
 \psfig{file=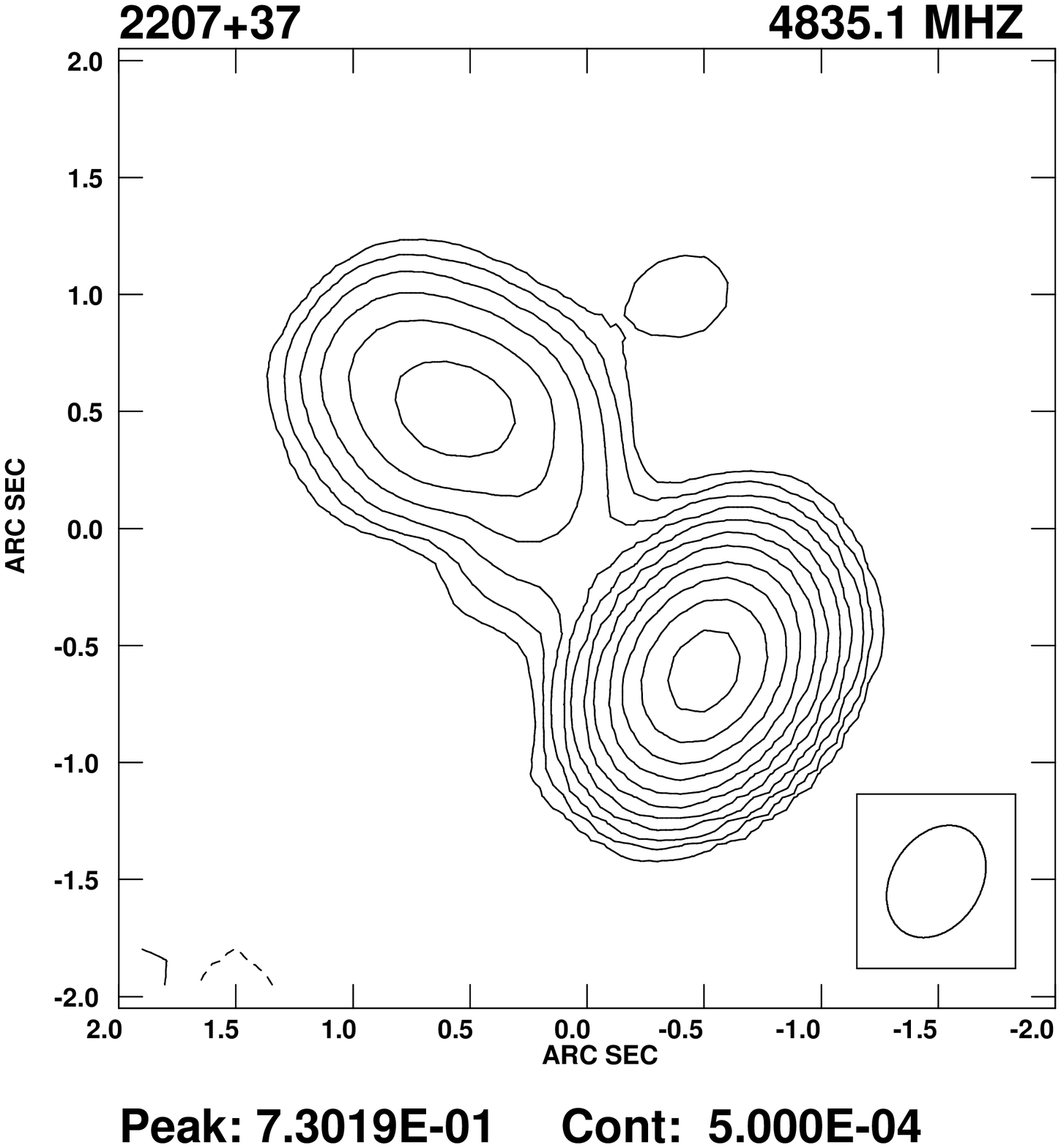,width=2.1in}
 }
\end{figure*}

\subsection{Notes on sources}

\noindent {\bf 0223+34:} There is evidence of weak emission to the 
         east of the core, which was noted by Spencer
         et al. (1989) but not confirmed by Dallacasa et al.
         (1995). VLBI observations show a bright compact
         component and a weaker one separated by about 
         0.$^{\prime\prime}$5 (Dallacasa et al. 1995).

\noindent {\bf 0655+69:} This is a triple source (Patnaik et al. 1992).
         Evidence of the diffuse emission to the east is
         barely seen in the high-resolution image by Patnaik et al. 

\noindent {\bf 0827+37:} A plume of emission
         to the north and diffuse emission to the south  are seen on the
         VLA $\lambda$20cm image of Xu et al. (1995).

\noindent {\bf 0902+49:} Comparison of the core flux density at $\lambda$6cm
         with those presented by Stanghellini et al. (1990)
         shows that this has not varied significantly over a period of about 1 yr.
         VLBI observations show the core to be extended (Henstock et al. 1995).
         It appears to have a one-sided radio structure.

\noindent {\bf 0945+66:} A lower-resolution $\lambda$20cm image by Xu et al.
         (1995) shows the source to be extended along a similar PA.

\noindent {\bf 1014+39:} A lower-resolution $\lambda$20cm image has been
         published by Machalski \& Condon (1983).

\noindent {\bf 1138+59:} VLBI observations show two sets of components with an
         overall separation of $\sim$225 mas along a PA of
         $\sim$170$^\circ$ (Xu et al. 1995).

\noindent {\bf 1818+35:} VLA observations at $\lambda$20 and 6cm show this to be
         a double source, with only a single component being
         detected on VLBI scales (Taylor et al. 1996).

\noindent {\bf 2207+37:} VLBI observations show a strongly curved jet-like
         structure towards the west (Thakkar et al. 1995; Xu et al. 1995).

\noindent {\bf 2311+46:} The VLBI image has a prominent component with an 
         extension, and has been modelled to consist of three
         components with an overall separation of $\sim$1.7 mas
         (Xu et al. 1995).

\begin{table*}
\caption{Symmetry parameters of sample of sources}
\begin{tabular}{l l l l l l l l l l}
\\
IAU & Alt. & Id & z & size & $f_c$ & $\Delta$ & $r_D$ & $r_L$ \\
\\
 0107+315 &      3C34 &  G &  0.689 &       222 &  0.00088 &      2.8 &     1.03 &      1.3 \\
 0125+287 &      3C42 &  G &  0.395 &       102 &   0.0033 &        7 &     1.07 &     1.25 \\
 0307+444 &  4C+44.07 &  Q &  1.165 &      24.6 &     0.14 &      5.5 &     1.65 &      5.5 \\
 0428+205 &     OF247 &  G &  0.219 &      0.51 & 0.12     &      4.5 &     4.69 &     0.16 \\
 0707+689 &  4C+68.08 &  Q &  1.139 &       6.7 &    0.059 &        7 &     1.24 &     2.15 \\
 0725+147 &     3C181 &  Q &  1.382 &        36 &    0.009 &        3 &      2.8 &      1.9 \\
 0806+426 &     3C194 &  G &  1.184 &        80 & 0.0060   &       24 &     1.03 &     0.66 \\
 1007+417 &  4C+41.21 &  Q &  0.611 &       147 &     0.23 &       10 &      1.6 &     20.3 \\
 1030+585 &   3C244.1 &  G &  0.428 &       193 &   0.0018 &      4.5 &      1.1 &      1.3 \\
 1108+359 &     3C252 &  G &  1.105 &       321 &   0.0052 &      4.5 &      1.8 &        2 \\
 1213+538 &  4C+53.24 &  Q &  1.065 &       190 &   0.0069 &      4.5 &     1.55 &      0.9 \\
 1225+368 &        B2 &  Q &  1.975 &      0.35 &     0.04 &     10.5 &      1.4 &     14.8 \\
 1317+520 &  4C+52.27 &  Q &   1.06 &       150 &     0.56 &       20 &      1.9 &     0.55 \\
 1358+624 &  4C+62.22 &  G &  0.431 &     0.149 & 0.0091   &     10.5 &     2.92 &     2.26 \\
 1420+198 &     3C300 &  G &  0.272 &       269 &   0.0075 &      4.5 &      2.4 &     0.32 \\
 1529+357 &     3C320 &  G &  0.342 &        51 & 0.015    &       11 &     1.76 &     0.97 \\
 1547+215 &     3C324 &  G &  1.206 &      63.5 &   0.0002 &        4 &     1.16 &      2.6 \\
 1549+628 &     3C325 &  G &   0.86 &      83.7 &   0.0025 &      3.5 &     1.32 &     0.48 \\
 1609+660 &     3C330 &  G &   0.55 &       263 &  0.00033 &      2.5 &     1.07 &     0.33 \\
 1627+234 &     3C340 &  G &  0.775 &       223 &   0.0024 &      1.5 &     1.06 &     2.27 \\
 1819+396 &  4C+39.56 &  G &  0.798 &      2.85 & 0.10     &      5.5 &     2.75 &     0.18 \\
 1829+290 &  4C+29.56 &  G &  0.842 &      11.8 & 0.96     &       21 &     1.65 &     1.23 \\
 1939+605 &     3C401 &  G &  0.201 &      44.8 &    0.027 &       14 &      1.4 &     1.34 \\
 1943+546 &       TXS &  G &  0.263 &      0.11 &     0.04 &        5 &     1.81 &     0.41 \\
 2230+114 &  4C+11.69 &  Q &  1.037 &        14 & 0.98     &      4.6 &      1.7 &     2.25 \\
 2324+405 &     3C462 &  G &  0.394 &        85 &    0.042 &        3 &     1.19 &     0.58 \\
 2342+821 &           &  Q &  0.735 &      0.83 & 0.14     &       11 &     1.39 &    36.15 \\
 2356+436 &     3C470 &  G &  1.653 &       155 & 0.0011   &      3.5 &     1.38 &     0.12 \\

\end{tabular}
\end{table*}

\section{Sample of sources and their symmetry parameters}

We note that although we use 20 kpc as a working definition for the upper limit 
to the size
of a CSS, any adopted value will be somewhat ad hoc. The detailed structure
of a source will be influenced by the external environment which could
sometimes be asymmetric on significantly larger scales, as can be seen in 
the optical broad- and narrow-band images of many high-redshift galaxies. Bearing this in mind,
we have considered the symmetry parameters for well-observed complete samples
of sources and studied their variations with linear size, with particular emphasis
on the CSSs. We have considered well-studied samples with a high degree of completeness
in optical identifications and redshift measurements; any missing objects are unlikely
to affect the trends reported in this paper. 
We have chosen the 3CR complete sample (Laing, Riley \& Longair
1983) with a radio luminosity at 178 MHz $\geq$10$^{26}$ W Hz$^{-1}$ sr$^{-1}$.
A luminosity limit higher than the value for differentiating
between the two Fanaroff \& Riley (1974) classes was chosen to minimise the
contamination by border-line objects which could be distorted Fanaroff-Riley
class I objects. To improve the statistics for high-luminosity sources, we have also
considered the complete sample of S4 radio sources (Kapahi 1981; Stickel \&
K\"{u}hr 1994) with $\alpha_{\sim1400}^{5000}$
$\geq$0.5 and luminosity at 5000 MHz $\geq$ 5$\times$10$^{24}$ W Hz$^{-1}$
sr$^{-1}$. The luminosity cutoff at 5000 MHz is similar to that at 178 MHz for 
a source with a spectral index of 0.8. In addition, we have considered  
the 3C and Peacock \& Wall (1982, hereinafter referred to as PW) samples of CSSs 
(Spencer et al. 1989) satisfying the above luminosity cutoffs.  
All the sources in our sample were required to have a good candidate or confirmed 
radio core. This is important for a reliable estimation of the 
symmetry parameters, particularly for the small sources. We have also not considered 
the complex sources whose symmetry parameters are not easily identifiable.
In this paper, we re-examine the trends reported by S95 which was based almost
entirely on the high-luminosity 3CR radio sources. We note that since the
appearance of S95, more radio cores have been identified in both 
CSSs and larger sources.

We are left with a sample of 109 sources, of  which 61 are galaxies, 
and 48 quasars. The 4 broad-line radio galaxies in the sample have been included 
with the quasars. Of these 109, there are 27 CSSs, of which 13 are galaxies and
14 quasars. The sample considered by S95 consisted of 81 sources. The 28 
additional sources are listed in Table 2 which is arranged as
follows: Column 1: source name; column 2: an alternative name; column 3: optical
identification where G denotes a galaxy and Q
a quasar; column 4: redshift; column 5: the projected linear size 
in kpc; column 6: the fraction of the emission from
the core at an emitted frequency of 8 GHz, assuming spectral indices of 0 and
1 for the nuclear and extended emission; 
column 7: the misalignment angle $\Delta$ which is the
supplement of the angle formed
at the core by the outer hot-spots; column 8: the ratio, r$_D$, of
the separation from the nucleus of the farther component 
to that of the nearer one; column 9:
the ratio, r$_L$, of the integrated flux densities of the farther
and nearer components. Usually, these have been estimated 
from low-resolution images at about 5 GHz to minimize the effects of any 
missing flux density.  While estimating r$_D$ and $\Delta$, we have
used the brightest peak in each outer component. Extensions
which are fainter than these by more than about a factor of 3 have been
ignored. 

\begin{figure}
\hspace{1.0cm}
\begin{center}
\vbox{
\psfig{file=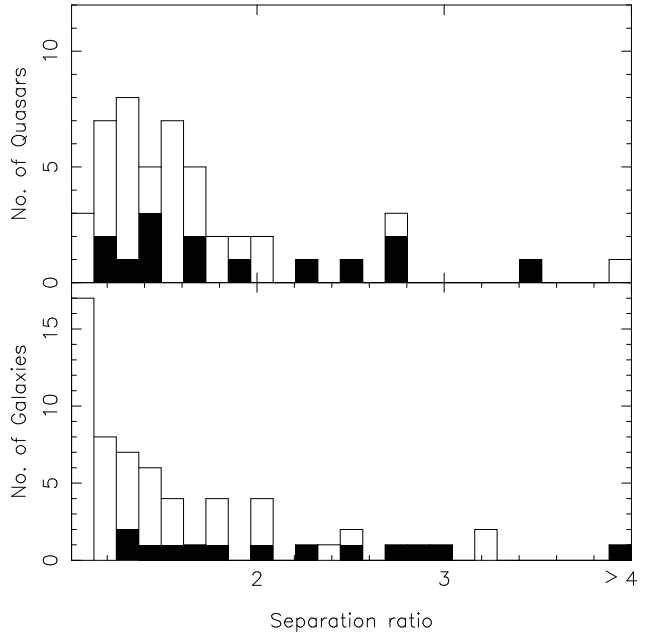,width=3.3in}
}
\end{center}
\caption{The distributions of the ratio, r$_D$, of the  separations of
the outer hot-spots from the nucleus, for the radio galaxies and quasars
in the entire sample. The CSS sources are shown in black.
}
\end{figure}

\section{Relativistic motion and symmetry parameters}
After including the results from our observations of the S4 sample, and more 
recent information from the literature, we have re-examined the trends reported 
in S95. We do not present most of the figures since the results
are similar to those presented in S95, but summarise the conclusions in this section.

\begin{enumerate}
\item The relative core strength for the quasars are larger than for the 
galaxies, consistent with the unified scheme. This is also true for the CSS 
subsamples, suggesting that their members are also consistent with the unified
scheme. The median values of f$_c$ are 0.07 and 0.004 for the quasars and
galaxies in the entire sample, and 0.10 and 0.007 for the CSS quasars
and galaxies. These values are consistent with the expectations of the
unified scheme (cf. Saikia \& Kulkarni 1994; Hardcastle et al. 1998).

\item The sources with strong cores, f$_c \geq$ 0.1, which are almost all 
quasars, and those with small sizes tend to be more misaligned. The median value
of $\Delta$ for sources with f$_c \geq$ 0.1 is 11$^\circ$ compared to about 
5.5$^\circ$ for the sources with weaker cores. Although this could be due to
projection effects, large misalignments in a number of CSSs with weak cores
suggests than interactions with the environment also play a significant role.

\item The distributions of the separation ratio, r$_D$, for the entire sample of
sources is shown in Figure 2. The CSSs are shown in black. The deficit of 
symmetric quasars, which is attributable to orientation, is clearly seen. It is also
striking that the CSSs associated with both radio galaxies and quasars have
a flatter distribution with higher median values. The median value of r$_D$
for the CSS galaxies is about 1.97 while the value for the larger galaxies is
1.33. The corresponding values for the quasars are 1.70 and 1.52. The smallest
value of r$_D$ for the CSS quasars and galaxies are about 1.17 and 1.33 
respectively.

\item The median flux density ratio, r$_L$, defined to be $>$1, for the CSS 
quasars and galaxies are
2.99 and 2.26 respectively, compared to 1.95 and 1.51 for the larger 
quasars and galaxies respectively. Although these appear to be consistent
with the trend expected in the unified scheme, many of the bright components
are closer to the nucleus, which is not as predicted (Figure 3). This is seen 
in a large fraction of
the CSSs, especially among those associated with galaxies where the effects
of orientation are minimal, and can be understood in terms of propagation of jets through an
asymmetric environment. The component on the denser side appears closer and 
also brighter due to higher dissipation of energy. Considering the most asymmetric 
objects in terms of the
location of the outer components, say r$_D \geq$ 1.8, 14 of the 15 galaxies have
the brighter component closer compared to 6 of the 11 quasars. 
Numerical and analytical estimates for the  propoagation
of jets through reasonably asymmetric environments on opposite sides of
the nucleus can reproduce the observed asymmetries (Jeyakumar et al., in preparation).
Evidence of such asymmetries in the distribution of gas, which might be
related to the fuelling of the radio source, is sometimes provided through a 
huge differential rotation measure on opposite sides of the nucleus
(Mantovani et al. 1994; Junor et al. 1999).

\end{enumerate}

\begin{figure}
\hspace{1.0cm}
\begin{center}
\psfig{file=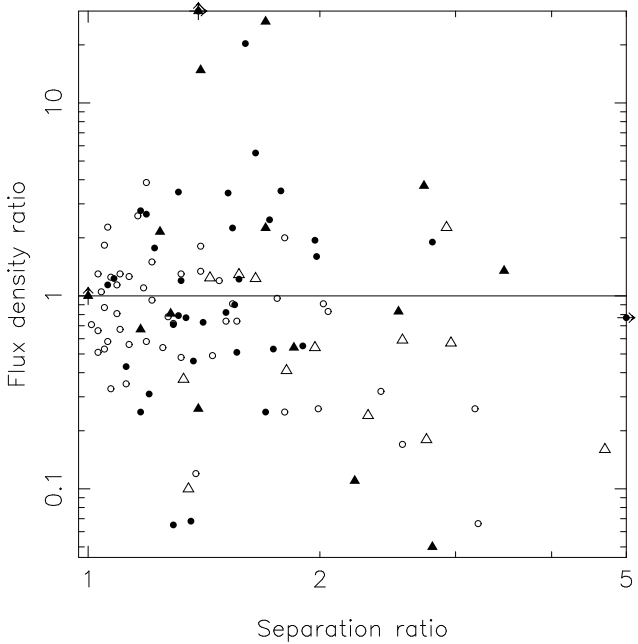,width=3.3in}
\end{center}
\caption{ The log r$_D$ - log r$_L$ diagram for the radio galaxies and
quasars. The filled and open triangles denote CSS quasars and radio galaxies, 
while filled and open circles denote larger-sized quasars and radio galaxies.
The sources whose values lie outside the range of the box shown are marked with an arrow.
}
\end{figure}

\section{Concluding remarks}
The radio properties of CSSs are consistent with the
unified scheme in which galaxies are seen close to the sky plane, while quasars
are inclined at small angles to the line of sight. However, the CSSs appear to be 
evolving in asymmetric environments. Comparison with larger sources
and theoretical and numerical estimates of the symmetry parameters as the jets 
propagate outwards through such an asymmetric environment (Jeyakumar et al., 
in preparation) suggest that most sources may have passed through such a phase.
These models yield ages for the CSSs in the range of $\sim$10$^5-10^6$ yr, 
consistent with estimates from emission line studies (de Vries et al. 1999) and
suggestions that the vast majority of CSSs are young sources
(Fanti et al. 1995; Readhead et al. 1996a,b). 

\section*{Acknowledgments}
We thank an anonymous referee for his valuable comments and suggestions.
The Very Large Array is operated by the National Radio Astronomy
Observatory for Associated Universities Inc. under a co-operative
agreement with the National Science Foundation.  The Arecibo
Observatory is part of the National Astronomy and Ionosphere Center,
which is operated by Cornell University under a cooperative agreement
with the National Science Foundation.
This research has made use of the NASA/IPAC extragalactic database (NED)
which is operated by the Jet Propulsion Laboratory, Caltech, under contract
with the National Aeronautics and Space Administration.

{}

\end{document}